\documentstyle[iopconf1]{article}

\def\be{\beta}
\def\ga{\gamma}
\def\de{\delta}
\def\ep{\epsilon}

\def\ka{\kappa}
\def\la{\lambda}

\def\si{\sigma}

\def\ph{\phi}

\def\ch{\chi}
\def\ps{\psi}

\def\Ga{\Gamma}
\def\De{\Delta}

\def\cl{{\cal L}}

\def\prt{\partial}
\def\vev#1{\langle {#1}\rangle}

\def\fr#1#2{{{#1} \over {#2}}}
\def\frac#1#2{{\textstyle{{#1}\over {#2}}}}
\def\half{{\textstyle{1\over 2}}}
\def\lsim{\mathrel{\rlap{\lower4pt\hbox{\hskip1pt$\sim$}}
    \raise1pt\hbox{$<$}}}
\def\gsim{\mathrel{\rlap{\lower4pt\hbox{\hskip1pt$\sim$}}
    \raise1pt\hbox{$>$}}}
\def\sqr#1#2{{\vcenter{\vbox{\hrule height.#2pt
         \hbox{\vrule width.#2pt height#1pt \kern#1pt
         \vrule width.#2pt}
         \hrule height.#2pt}}}}

\def\lrprtmu{\stackrel{\leftrightarrow}{\partial_\mu}}
\def\lrprtnu{\stackrel{\leftrightarrow}{\partial^\nu}}

\def\Im{\hbox{Im}\,}

\newcommand{\beq}{\begin{equation}}
\newcommand{\eeq}{\end{equation}}
\newcommand{\bea}{\begin{eqnarray}}
\newcommand{\eea}{\end{eqnarray}}
\newcommand{\rf}[1]{(\ref{#1})}

\begin{document}

\begin{flushright}
IUHET 413\\
August 1999
\end{flushright}
\vglue 0.5 cm

\title{Lorentz- and CPT-Violating Extension\\
of the Standard Model\footnote{Invited talk at the conference 
\it Beyond the Standard Model, \rm
Tegernsee, Germany, June 1999}
}

\author{V.\ Alan Kosteleck\'y}

\affil{Physics Department, Indiana University,\\
Bloomington, IN 47405, U.S.A.}

\beginabstract
The formulation and some experimental implications
of a general Lorentz-violating extension of the
standard model are reviewed.
The theory incorporates both CPT-preserving
and CPT-breaking terms.
It is otherwise a conventional quantum field theory,
obtained under the assumption that Lorentz symmetry
is spontaneously broken in an underlying model.
The theory contains the usual standard-model gauge structure,
and it is power-counting renormalizable.
Energy and momentum are conserved.
Despite the violation of Lorentz symmetry,
the theory exhibits covariance under Lorentz transformations
of the observer inertial frame.
A general Lorentz-violating extension of quantum electrodynamics
can be extracted.
The standard-model extension implies potentially observable effects
in a wide variety of experiments,
including among others
measurements on neutral-meson oscillations,
comparative studies in Penning traps,
spectroscopy of hydrogen and antihydrogen,
bounds on cosmological birefringence,
measurements of muon properties,
clock-comparison tests,
and observations of the baryon asymmetry.
\endabstract

\section{Introduction}

The standard model of particle physics
is invariant under the discrete operation CPT,
which combines charge conjugation C,
parity inversion P, and time reversal T.
Indeed,
the product CPT is the only remaining combination of C, P, T
generally believed to be an exact symmetry of nature.
The standard model also exhibits symmetry under
continuous Lorentz transformations,
which include both rotations and boosts.
Both CPT and Lorentz symmetries are connected
via the CPT theorem,
which states under mild technical assumptions
that CPT is an exact symmetry of local Lorentz-covariant field theories 
of point particles \cite{cpt,sachs}.

Other than theoretical prejudice,
there is strong support from high-precision experiments
in favor of Lorentz and CPT symmetry of nature.
For example,
the Particle Data Group identifies 
\cite{pdg}
the sharpest figure of merit for CPT tests
as one involving the kaon particle-antiparticle mass difference,
currently constrained by experiments at Fermilab and CERN
to \cite{kexpt}
\beq
\fr{|m_K - m_{\overline K}|}{m_K}
\lsim 10^{-18}
\quad .
\label{a}
\eeq
Tight constraints on possible Lorentz violations also exist.

The presence of 
a general CPT theorem for Lorentz-covariant particle theories 
and the establishment of high-precision experimental tests 
implies that the observation of Lorentz or CPT violation
would represent a sensitive signal for unconventional physics
beyond the standard model.
Possible theoretical mechanisms 
through which Lorentz or CPT symmetry might be violated
are therefore worth investigating
\cite{cpt98}. 
However,
straightforward approaches typically 
require radical revisions of conventional quantum field theory
or contain physical features such as infinite-spin particles
that seem unlikely to be realized in nature. 

Despite this, 
at least one promising theoretical possibility exists 
that allows Lorentz and CPT violation in the context of the standard model.
It is based on the idea of spontaneous breaking 
of Lorentz and CPT symmetry
in an underlying theory \cite{ks,kp},
and it appears to be compatible both with experiment
and with established quantum field theory.
This theory leads to a general phenomenology 
for Lorentz and CPT violation
at the level of the standard model and quantum electrodynamics (QED)
\cite{kp2,cksm}.

The resulting standard-model extension
indicates that apparent breaking of Lorentz and CPT symmetry
could be observable in a variety of existing or feasible experiments.
Relatively few experiments testing Lorentz and CPT symmetry 
have the necessary sensitivity to observe possible signals.
However,
a few high-precision ones already bound 
some of the parameters for Lorentz and CPT violation 
in the standard-model extension.
Among the ones investigated to date are experiments with 
neutral-meson oscillations \cite{kexpt,kp,kp2,ckv,bexpt,ak},
comparative tests of QED 
in Penning traps \cite{pennexpts,bkr,gg,hd,rm},
spectroscopy of hydrogen and antihydrogen \cite{antih,bkr2},
measurements of muon properties
\cite{bkl},
clock-comparison experiments
\cite{cc,kla},
measurements of cosmological birefringence \cite{cfj,cksm,jk,pvc},
and observations of the baryon asymmetry \cite{bckp}.
Effects on cosmic rays have also been investigated 
in a restricted version of the standard-model extension
\cite{cg}.

In the remainder of this talk,
the formulation and experimental implications
of the general Lorentz- and CPT-breaking standard-model extension 
are briefly described.
Some different approaches to unconventional spacetime physics 
beyond the standard model are discussed in other presentations
at this meeting
\cite{ra,nm,dn,hp}.

\section{Theoretical framework}

Developing a 
satisfactory microscopic theory allowing Lorentz and CPT breaking
is a difficult task.
It is therefore tempting to avoid the issue 
via a purely phenomenological approach,
in which one introduces a parametrization of  
observable quantities
that allows for Lorentz or CPT violation.
An illustration is provided by the phenomenology of CPT violation 
in neutral-kaon oscillations \cite{leewu}.
In this case,
the physical kaon eigenstates $K_S$ and $K_L$
are formed as linear combinations of the strong-interaction eigenstates
$K^0$ and $\overline{K^0}$
involving two complex quantities 
that parametrize CP violation.
One quantity, $\ep_K$, parametrizes T violation with CPT symmetry 
while the other, $\de_K$, parametrizes CPT violation with T symmetry.
In the usual standard model,
a mechanism for T violation exists 
so that a nonzero value of $\ep_K$ is calculable in principle.
In contrast, 
CPT is a symmetry of the usual standard model 
and $\de_K$ therefore vanishes.

Introducing the parameter $\de_K$
is in this context a purely phenomenological choice,
without basis in a microscopic theory.
In fact,
it is unclear \it a priori \rm whether this parametrization 
can make physical sense in a microscopic theory.
Certainly,
without a microscopic description
$\de_K$ cannot be related to different
phenomenological parameters for CPT violation in other systems.
Similar difficulties face other phenomenological descriptions
of Lorentz and CPT violation.
It would therefore seem more desirable 
to develop an explicit microscopic theory 
for such effects.
Ideally,
one would like a theoretical framework
at the level of the standard model
from which one could extract a quantitative and general 
phenomenology for Lorentz and CPT violation
in any system of interest.
If available,
such a theory would permit 
direct comparisons between experiments,
calculation of phenomenological parameters,
and possibly the prediction of signals.

A microscopic theory of this type
can be obtained within the framework  
of spontaneous Lorentz and CPT breaking \cite{ks,kp}.
The basic notion is that 
apparent violation of Lorentz and CPT symmetries
could emerge from spontanteous symmetry breaking 
in a fully Lorentz- and CPT-covariant fundamental theory of nature. 
This mechanism appears theoretically viable 
and is therefore a relatively elegant way to 
break Lorentz and CPT invariance.
Spontaneous breaking has the advantage 
that many desirable features of 
a Lorentz-covariant theory can be anticipated
because the dynamics remains covariant,
even though the solutions 
and the physics of the resulting theory 
exhibit Lorentz and CPT violations.
In contrast,
other types of Lorentz and CPT breaking are typically
inconsistent with theoretical notions such as 
probability conservation or microcausality.

In some respects,
the behavior of a particle
in a vacuum with spontaneous Lorentz and CPT violation
is related to that of a conventional particle 
moving inside a biaxial crystal \cite{cksm}.
The presence of the crystal typically breaks
both rotational and boost symmetry,
but these features are compatible with 
a consistent theoretical description
and properties such as causality.
Instead of being associated with fundamental problems,
the lack of Lorentz covariance here
merely follows from the existence 
of the background crystal fields.
In fact,
microcausality can explicitly be demonstrated
in some simple models emerging  
from spontaneous Lorentz and CPT breaking \cite{cksm}.

Spontaneous breaking of Lorentz and CPT symmetry
can be triggered in a Lorentz-covariant theory
by certain types of interaction among Lorentz-tensor fields
that destabilize the naive vacuum 
and lead to nonzero Lorentz-tensor expectation values.
The true vacuum fills with
these tensor expectation values,
producing spontaneous Lorentz breaking \cite{ks},
together with spontaneous CPT breaking 
if the tensor expectation values
have an odd number of spacetime indices \cite{kp}.
Apparent violations of Lorentz and CPT symmetry 
could arise at the level of the standard model \cite{kp2}
if some components of the tensor expectation values lie 
along the four macroscopic spacetime dimensions.
Some of the possible observable effects in experiments
that could result from this are discussed below.

The destabilizing Lorentz-tensor interactions
required to trigger spontaneous Lorentz violation 
are absent in conventional four-dimensional renormalizable gauge theories,
including the usual standard model.
Suitable tensor interactions do occur,
however,
in some string (M) theories.
The issue of realizing spontaneous Lorentz violation
in this context can be investigated,
for example,
using string field theory
in the particular case of the open bosonic string.
For this case,
the action and equations of motion
for particle fields below a specified level number $N$
can be derived analytically.
One can then obtain the associated solutions 
and identify those that persist as $N$ increases
\cite{ks}.
This procedure has been completed for some cases 
to a depth of over 20,000 terms in 
the static potential \cite{kp}.
Solutions that spontaneously break Lorentz symmetry
are among those remaining stable as $N$ increases.

When a continuous global symmetry
is spontaneously broken
in a conventional field theory,
corresponding massless modes arise
in accordance with 
the Nambu-Goldstone theorem.
Similarly,
spontaneous breaking of a continuous global Lorentz symmetry
would also lead to massless modes.
In contrast,
when a global spontaneously broken symmetry
is promoted to a local gauge symmetry
the Higgs mechanism occurs,
in which the massless modes disappear
and a mass is generated for the gauge boson. 
However,
although the inclusion of gravity
promotes Lorentz invariance to a local symmetry,
there is no analogue to the Higgs effect \cite{ks}
because connection depends on
derivatives of the metric rather than directly on the metric itself.
Thus,
when local Lorentz symmetry is spontaneously broken,
the graviton propagator is modified
but no graviton mass is generated.

\section{Standard-Model Extension and QED Limit}

For consistency with known experimental limits,
any apparent breaking of Lorentz and CPT symmetries 
occurring at the level of the SU(3)$\times$SU(2)$\times$U(1) standard model
and QED must be highly suppressed.
If indeed this originates 
in spontaneous breaking in a Planck-scale fundamental theory,
then the natural dimensionless suppression factor 
is the ratio of a low-energy scale associated with the
standard model to the Planck scale.
This heavy suppression implies that only limited effects 
of Lorentz or CPT violation would be observable.

At low energies,
effects from any apparent Lorentz and CPT violation
would be governed by a standard-model extension
that arises as the low-energy limit 
of the fundamental theory \cite{kp2}.
Intuition about the structure of the standard-model extension
can be obtained by considering various possible couplings
in the low-energy limit of the fundamental theory
and examining the resulting form when Lorentz tensors
acquire expectation values.
For instance,
possible trilinear couplings between fermions and 
one or more boson tensor fields can produce 
terms in the low-energy theory of the form
\beq
\cl \sim \fr {\la} {M^k} 
\vev{T}\cdot\overline{\ps}\Ga(i\prt )^k\ch
+ {\textstyle h.c.}
\quad ,
\label{aa}
\eeq
where the tensor expectation values are denoted $\vev{T}$.
The Lorentz properties of the 
bilinear in the fermion fields $\ps$, $\ch$
are fixed by the gamma-matrix structure $\Ga$
and the $k$ spacetime derivatives $i\prt$,
and these properties establish 
the associated apparent Lorentz and CPT violation 
in the low-energy theory.
In addition to the expectation value $\vev{T}$,
the coefficient of the Lorentz-breaking fermion bilinear
involves a dimensionless coupling $\la$
and an appropriate power of some large 
(compactification or Planck) scale $M$.

General considerations along the above lines
make it possible to establish all terms
at the level of the standard model
that are compatible with an origin in
spontaneous Lorentz and CPT breaking,
no matter what the form of the fundamental theory.
Imposing also the requirements of
SU(3)$\times$SU(2)$\times$U(1) gauge invariance
and power-counting renormalizability
produces a general hermitian standard-model extension 
allowing for Lorentz violation
with both CPT-even and CPT-odd terms \cite{cksm}.
This theory at present appears to be the only candidate
for a consistent standard-model extension
originating in a microscopic theory of Lorentz violation.

Since the apparent Lorentz and CPT violation
arises from spontaneous breaking in a fully Lorentz-covariant theory,
the resulting standard-model extension 
has a number of attractive features
common to the usual Lorentz-covariant field theories
\cite{cksm}.
For example,
standard quantization methods can be used,
and properties such as positivity of the energy
and microcausality can be expected to hold. 
Assuming the tensor expectation values 
are independent of spacetime location
(no soliton solutions),
then energy and momentum are conserved.
It can also be shown that the usual  
gauge-symmetry breaking to the electromagnetic U(1) occurs,
so the usual gauge structure is unaffected.
Moreover,
the theory is 
covariant under observer Lorentz transformations:
rotations or boosts of the observer's inertial frame
leave unaffected the physics because 
the background expectation values
transform covariantly along with the localized fields.
Instead,
the apparent Lorentz violations are associated with
particle Lorentz transformations:
rotations or boosts of the localized fields 
in a fixed observer inertial frame 
can change the physics because
they leave untouched the background expectation values.

The usual forms of QED can be extracted as suitable limits
from the conventional standard model.
Similarly,
Lorentz-violating QED extensions
emerge by taking
suitable limits of the standard-model extension
\cite{cksm}.
Both the fermion and the photon sectors
acquire apparent Lorentz-violating terms
with CPT-even and CPT-odd contributions.
Since there exist many high-precision experiments with QED,
the Lorentz-violating QED extensions are of
particular interest.

The simplest explicit example
is the extension of QED that
involves only electrons, positrons, and photons.
The usual lagrangian is:
\beq
\cl^{\rm QED} =
\overline{\ps} \ga^\mu (\half i \lrprtmu - q A_\mu ) \ps 
- m \overline{\ps} \ps 
- \frac 1 4 F_{\mu\nu}F^{\mu\nu}
\quad .
\label{aaa}
\eeq
The limiting QED case from the standard-model extension
produces CPT-odd terms,
\beq
\cl^{\rm CPT}_{e} =
- a_{\mu} \overline{\ps} \ga^{\mu} \ps 
- b_{\mu} \overline{\ps} \ga_5 \ga^{\mu} \ps \quad ,
$$
$$
\cl^{\rm CPT}_{\ga} =
\half (k_{AF})^\ka \ep_{\ka\la\mu\nu} A^\la F^{\mu\nu}
\quad ,
\label{bbb}
\eeq
and CPT-even ones,
\beq
\cl^{\rm Lorentz}_{e} = 
c_{\mu\nu} \overline{\ps} \ga^{\mu} 
(\half i \lrprtnu - q A^\nu ) \ps 
+ d_{\mu\nu} \overline{\ps} \ga_5 \ga^\mu 
(\half i \lrprtnu - q A^\nu ) \ps 
- \half H_{\mu\nu} \overline{\ps} \si^{\mu\nu} \ps 
$$
$$
\cl^{\rm Lorentz}_{\ga} =
-\frac 1 4 (k_F)_{\ka\la\mu\nu} F^{\ka\la}F^{\mu\nu}
\quad .
\label{ccc}
\eeq
All coefficients of the extra terms 
can be regarded as minuscule Lorentz- and CPT-violating couplings.
All the extra terms are invariant under
observer Lorentz transformations
but break particle Lorentz covariance.
Note that some of the coefficient components 
are unobservable physically,
because field redefinitions must be taken into account.
For instance,
coefficients of the type $a_\mu$ can be directly detected only 
in flavor-changing experiments,
and so at leading order they are unobservable 
in experiments restricted to electrons, positrons, and photons.
Ref.\ \cite{cksm}
provides additional information about the properties 
of the above expressions,
including details of the notation and conventions.

\section{Experiment}

A variety of high-precision experimental tests of 
Lorentz and CPT symmetry
can be quantitatively investigated and compared
within the standard-model extension described above,
and in favorable cases potentially observable signals
can be identified.
The remainder of this talk 
provides a short overview of 
some of the results obtained to date.

\subsection{Neutral-Meson Oscillations}

Four neutral-meson systems are known to exhibit flavor oscillations:
$K$, $D$, $B_d$, and $B_s$.
In what follows,
a generic neutral meson is denoted by $P$.
The time evolution of 
a neutral-meson state is governed by 
a non-hermitian two-by-two effective hamiltonian
in the meson-antimeson state space.
Two complex parameters controlling indirect CP violation
appear in the effective hamiltonian:
$\ep_P$ and $\de_P$.
The quantity $\ep_P$ parametrizes T violation,
while $\de_P$ parametrizes CPT violation.
For the $K$ system,
$\ep_K$ and $\de_K$
are the phenomenological quantities 
discussed in section 2.
Bounds on the magnitude of $\de_P$
provide constraints on CPT violation 
in the neutral-meson systems.

The quantity $\de_P$ vanishes 
in the context of the usual standard model
because CPT is a symmetry.
However,
$\de_P$ can be derived 
in the context of the standard-model extension
\cite{ak}.
At leading order,
it can be shown that $\de_P$ varies with
only one kind of Lorentz- and CPT-violating term 
in the standard-model extension,
of the form
$- a^q_{\mu} \overline{q} \ga^\mu q$.
Here, 
$q$ represents a valence quark field in the $P$ meson
and the quantity $a^q_{\mu}$
varies with quark flavor $q$ but is spacetime constant.
An expression for $\de_P$ 
in the frame in which the quantities $a^q_{\mu}$ are specified,
valid at leading order in all coupling coefficients 
in the standard-model extension,
is \cite{ak}
\beq
\de_P \approx i \sin\hat\ph \exp(i\hat\ph) 
\ga(\De a_0 - \vec \be \cdot \De \vec a) /\De m
\quad .
\label{e}
\eeq
Here,
$\be^\mu \equiv \ga(1,\vec\be)$ 
denotes the four-velocity of the $P$-meson.
Its appearance in the expression for $\de_P$ 
represents a variation with the boost and orientation of the $P$ meson
and is a reflection of the breaking of Lorentz symmetry 
in the standard-model extension.
In Eq.\ \rf{e},
$\De a_\mu \equiv a_\mu^{q_2} - a_\mu^{q_1}$,
where $q_1$ and $q_2$ are the valence-quark flavors 
for the $P$ meson,
and $\hat\ph\equiv \tan^{-1}(2\De m/\De\ga)$,
where $\De m$ and $\De \ga$
are the mass and decay-rate differences,
respectively,
between the $P$-meson eigenstates.
Note that subscripts $P$ have been omitted for simplicity.

The result \rf{e} implies several interesting features
of tests of Lorentz and CPT symmetry with neutral mesons
and makes several predictions.
For example,
the dependence on $a^q_\mu$ and the independence of
other coefficients for Lorentz violation
in the standard-model extension,
together with the involvement of flavor-changing effects,
means that neutral-meson tests are independent at leading order of 
any results from other experiments discussed in this talk.
Another point is that 
the real and imaginary parts of $\de_P$ 
are predicted to be proportional \cite{kp2}.

Equation \rf{e} also suggests
that the magnitude of $\de_P$ 
may differ for distinct $P$ 
as a result of the flavor dependence 
of the coefficients $a_\mu^q$.
One possible scenario would be that 
the coefficients $a_\mu^q$ grow with mass,
as do the usual Yukawa couplings,
in which case the heavier neutral mesons
such as $D$ or $B_d$ may exhibit 
large CPT-violating effects.
Other interesting effects are predicted from
the dependence of the result \rf{e}
on the meson boost magnitude and orientation
\cite{ak}.
For instance,
distinct experiments may have inequivalent 
Lorentz and CPT reach
even if the statistical sensitivity is comparable.
This could occur if the mesons in one experiment
have a $4\pi$ distribution
while those for another are well collimated,
or if the meson-momentum spectra differ greatly 
in the two experiments.
Another possibility would be  
variations of the data with sidereal time,
arising from the rotation of the Earth relative to 
the orientation of the coefficients for Lorentz violation \cite{ak}.
Since the data in neutral-meson experiments are typically 
taken over an extended time period,
the consequences of time averaging must be taken into account
to provide a complete analysis of CPT violation.

Observations of the neutral-$K$ system
currently provide the sharpest clean experimental constraints
on CPT violation.
Both time-averaged limits on $\de_K$ and 
limits on the amplitude of sidereal variations
are now available \cite{kexpt}.
The heavier neutral-meson systems 
have also been used to place some
CPT bounds.
Two collaborations \cite{bexpt}
at CERN have studied the issue of whether \cite{ckv}
existing data suffice to constrain CPT violation.
The OPAL collaboration has published the measurement 
$\Im\de_{B_d} = -0.020 \pm 0.016 \pm 0.006$,
while the DELPHI collaboration has announced a preliminary 
result of $\Im\de_{B_d} = -0.011 \pm 0.017 \pm 0.005$.
Additional theoretical and experimental studies are 
in progress for all these systems.

\subsection{QED Experiments}

A practical approach to obtaining high-precision comparative data
for particles and antiparticles is to trap them individually
for extended time periods.
Experiments of this type can yield sharp CPT bounds
and thereby constrain the coefficients for Lorentz violation
in the fermion sector of the QED extension \cite{bkr}.
One option is to use a Penning trap to perform 
comparative studies of anomaly and cyclotron frequencies 
for particles and antiparticles
\cite{pennexpts}.
Several effects are predicted by the QED exension,
including direct signals in the form of frequency shifts
and variations of frequencies with sidereal time
as the Earth rotates \cite{bkr}.
Various experimental analyses along these lines have been performed,
and for a given experimental scenario 
one can estimate the attainable sensitivity
and introduce appropriate figures of merit for the
different effects.

An immediate CPT test is provided by
comparisons of the anomalous magnetic moments
of electrons and positrons.
This generates a constraint
on the spatial components of the coefficient $b^e_\mu$
in the laboratory frame.
A recent reanalysis of data from earlier experiments
has placed a bound of $1.2 \times 10^{-21}$
on the associated figure of merit
\cite{hd}.
Another test, 
which requires measurements only on the electron anomaly frequency,
involves searching for frequency variations with sidereal time.
A new experimental result
with a figure of merit bounded at $6\times 10^{-21}$
has recently been obtained
\cite{rm}.
Experiments similar to the above but performed with protons and antiprotons 
would be of interest.

Comparisons of particle and antiparticle cyclotron frequencies 
are also possible.
An elegant recent experiment
determines the cyclotron frequencies of antiprotons
and $H^-$ ions caught in the same trap \cite{gg}.
In the context of the standard-model extension,
leading-order effects in this experiment
are sensitive to Lorentz violation,
and the associated figure of merit 
is constrained to $4\times 10^{-25}$.

The possibility also exists of testing 
Lorentz and CPT symmetry 
to high precision by comparing spectroscopic data 
from trapped hydrogen and antihydrogen \cite{antih,bkr2}.
Various experimental signals can be considered.
Within the context of the standard-model and QED extensions,
certain 1S-2S and hyperfine transitions 
in magnetically trapped hydrogen and antihydrogen
provide direct sensitivity
to parameters for Lorentz and CPT violation.
For some specific transitions,
theoretically clean signals
for Lorentz and CPT violation exist 
that are unsuppressed by powers of the fine-structure constant.

Clock-comparison experiments 
\cite{cc}
provide an exceptionally 
sensitive means of searching for Lorentz violation.
Typically,
they constrain possible spatial anisotropies
by bounding the relative change 
of two hyperfine or Zeeman transition frequencies
as the Earth rotates.
Limits from experiments already performed
have recently been analyzed in the context of 
the standard-model extension
\cite{kla}.
The possible signals depend on the species of atom
used as the clocks.
The resulting experimental effects are controlled 
by a combination of proton, neutron, and electron parameters 
for Lorentz and CPT violation,
with bounds on suitable combinations of parameters
ranging from $10^{-25}$ to $10^{-30}$ GeV
under certain simplifying assumptions.

In the photon sector,
combining theoretical factors with 
terrestrial, astrophysical,
and cosmological experiments on electromagnetic radiation
provides sharp bounds on Lorentz-violating terms in
the QED extension.
The pure-photon CPT-violating term in Eq.\ (4) 
is known to contribute negatively 
to the energy \cite{cfj},
which would appear to restrict its viability
and indicate that the coefficient
$(k_{AF})^\ka$ should vanish \cite{cksm}.
This theoretical difficulty is absent from 
the CPT-even term in the following equation,
which is known to maintain a positive conserved energy \cite{cksm}.

Solving the extended Maxwell equations 
in the presence of Lorentz- and CPT-breaking effects
generates two independent propagating degrees of freedom \cite{cksm},
in agreement with the conventional case.
However,
in the extended Maxwell case
there are different dispersion relations for the two modes,
an effect that differs qualitatively from 
the usual propagation of electromagnetic waves in vacuum.
In fact,
in the presence of Lorentz and CPT violation,
an electromagnetic wave traveling in the vacuum
exhibits effects closely analogous to those displayed by 
an electromagnetic wave in conventional electrodynamics 
that is passing through 
a transparent optically anisotropic and gyrotropic crystal 
with spatial dispersion of the axes \cite{cksm}.
This effect leads to sharp experimental limits
on the coefficients for Lorentz violation 
in the extended Maxwell theory,
extracted from bounds on the birefringence of radio waves 
on cosmological distance scales.
For the CPT-odd coefficient $(k_{AF})_\mu$,
present constraints on cosmological birefringence
place a limit of the order of $\lsim 10^{-42}$ GeV
on its components
\cite{cfj,hpk}.
For the CPT-even dimensionless coefficient $(k_F)_{\ka\la\mu\nu}$,
the single rotation-invariant irreducible component 
is constrained to $\lsim 10^{-23}$ 
by the existence of cosmic rays \cite{cg}
and other tests.
All other irreducible components
of $(k_F)_{\ka\la\mu\nu}$
are associated with violations of rotation invariance,
and in principle it should be possible to bound 
these coefficients 
at the level of about $10^{-27}$
with existing methods for
measuring cosmological birefringence \cite{cksm}.
However,
no such bounds presently exist.

The theoretically favored zero value of 
the coefficient $(k_{AF})_\mu$ 
is compatible with
the sharp experimental constraints obtained from
the absence of cosmological birefringence.
However,
a zero tree-level value of $(k_{AF})_\mu$
is not protected by any symmetry,
and so one might plausibly anticipate 
that $(k_{AF})_\mu$ would acquire a finite contribution 
from radiative corrections involving CPT-violating couplings
in the fermion sector.
Remarkably,
this difficulty can be avoided
by an anomaly-cancellation mechanism,
which ensures the finiteness of the net sum 
of all one-loop radiative corrections.
Although the contribution from each individual radiative correction
is ambiguous \cite{cksm,jk,pvc},
the anomaly-cancellation mechanism can be applied
even if the theory is defined
such that each individual radiative correction is nonzero.
A tree-level CPT-odd term is therefore unnecessary
for one-loop renormalizability.
Related cancellations may occur at higher loops.
The freedom to select a zero tree-level value 
for a CPT-odd term that appears unprotected by 
any symmetry represents a significant theoretical test 
of the internal consistency of the standard-model extension.
In contrast, 
it can be shown that no such mechanism exists
for the CPT-even Lorentz-violating pure-photon term
\cite{cksm}.
Explicit calculation at the one-loop level
demonstrates the existence of divergent radiative corrections.
Future observation of a nonzero cosmological birefringence
asssociated with this term therefore remains an open possibility.

A variety of other potentially observable Lorentz and CPT signals
are known.
For instance,
under suitable conditions
the observed baryon asymmetry can be produced in thermal equilibrium
through Lorentz- and CPT-violating bilinear terms \cite{bckp}.
In the usual picture without CPT violation, 
nonequilibrium processes
and C- and CP-breaking interactions are needed \cite{ads}.
In contrast,
the unconventional scenario with CPT violation
can generate a relatively large baryon asymmetry at grand-unified scales
that subsequently dilutes
to the observed value through sphaleron or other effects.

\section*{Acknowledgments}
I thank Orfeu Bertolami, Robert Bluhm, Chuck Lane,
Don Colladay, Roman Jackiw, Malcolm Perry,
Rob Potting, Neil Russell, Stuart Samuel, 
and Rick Van Kooten for collaborations,
and Hans Klapdor-Kleingrothaus for hospitality
and organization of a stimulating meeting.
The research described in this paper was supported in part
by the United States Department of Energy 
under grant number DE-FG02-91ER40661,
by the North Atlantic Treaty Organization under
grant number CRG 960693,
and by the F.C.T. (Portugal).


\begin{thebibliography}{99}

\bibitem{cpt}
J.S. Bell, Birmingham University thesis (1954);
G.\ L\"uders,
Det.\ Kong.\ Danske Videnskabernes Selskab Mat.\-fysiske
Meddelelser 28, no.\ 5 (1954);
W.\ Pauli,
in W.\ Pauli, ed.,
\it Niels Bohr and the Development of Physics \rm
(McGraw-Hill, New York, 1955).

\bibitem{sachs}
For a textbook discussing CPT,
see, for example, R.G.\ Sachs,
{\it The Physics of Time Reversal}
(University of Chicago Press, Chicago, 1987).

\bibitem{pdg}
See, e.g., 
Review of Particle Properties,
Eur.\ Phys.\ J.\ C {\bf 3} (1998) 1.

\bibitem{kexpt}
B. Schwingenheuer 
{\it et al.},
Phys.\ Rev.\ Lett.\ {\bf 74} (1995) 4376;
L.K. Gibbons et al.,
Phys. Rev. D {\bf 55} (1997) 6625;
R. Carosi et al., Phys. Lett. B {\bf 237} (1990) 303;
KTeV Collaboration,
presented by Y.B.\ Hsiung at the KAON 99 conference, Chicago (1999);
CPLEAR Collaboration,
presented by P.\ Bloch at the KAON 99 conference, Chicago (1999).

\bibitem{cpt98}
Reviews of various theoretical and experimental 
approaches can be found in, 
for example, 
V.A.\ Kosteleck\'y, ed.,
{\it CPT and Lorentz Symmetry}
(World Scientific, Singapore, 1999).

\bibitem{ks}
V.A.\ Kosteleck\'y and S.\ Samuel,
Phys.\ Rev.\ Lett.\ {\bf 63} (1989) 224;
{\it ibid.},
{\bf 66} (1991) 1811;
Phys.\ Rev. D {\bf 39} (1989) 683;
{\it ibid.},
{\bf 40} (1989) 1886.

\bibitem{kp}
V.A.\ Kosteleck\'y and R.\ Potting,
Nucl.\ Phys.\ B {\bf 359} (1991) 545;
Phys.\ Lett.\ B {\bf 381} (1996) 89;
V.A.\ Kosteleck\'y, M.J.\ Perry, and R.\ Potting,
IUHET preprint 412 (1999), hep-th/9912243.

\bibitem{kp2}
V.A.\ Kosteleck\'y and R.\ Potting,
Phys.\ Rev.\ D {\bf 51} (1995) 3923;
and in D.B.\ Cline, ed.,
{\it Gamma Ray--Neutrino Cosmology and Planck Scale Physics} \rm
(World Scientific, Singapore, 1993)
(hep-th/9211116).

\bibitem{cksm}
D.\ Colladay and V.A.\ Kosteleck\'y,
Phys.\ Rev.\ D {\bf 55} (1997) 6760;
Phys.\ Rev.\ D {\bf 58} (1998) 116002.

\bibitem{ckv}
D.\ Colladay and V. A. Kosteleck\'y,
Phys.\ Lett.\ B {\bf 344} (1995) 259;
Phys.\ Rev.\ D {\bf 52} (1995) 6224;
V.A.\ Kosteleck\'y and R.\ Van Kooten,
Phys.\ Rev. D {\bf 54} (1996) 5585.

\bibitem{bexpt}
OPAL Collaboration, 
R.\ Ackerstaff
{\it et al.},
Z.\ Phys. C {\bf 76} (1997) 401;
DELPHI Collaboration,
M.\ Feindt
{\it et al.},
preprint DELPHI 97-98 CONF 80 (July 1997).

\bibitem{ak}
V.A.\ Kosteleck\'y,
Phys.\ Rev.\ Lett.\ {\bf 80} (1998) 1818;
Phys.\ Rev. D {\bf 61} (2000) 16002.

\bibitem{pennexpts}
P.B.\ Schwinberg, R.S.\ Van Dyck, Jr., and H.G.\ Dehmelt,
Phys.\ Lett.\ A {\bf 81} (1981) 119;
Phys.\ Rev.\ D {\bf 34} (1986) 722;
L.S.\ Brown and G.\ Gabrielse,
Rev.\ Mod.\ Phys.\ {\bf 58} (1986) 233;
R.S.\ Van Dyck, Jr., P.B.\ Schwinberg, and H.G.\ Dehmelt, 
Phys.\ Rev.\ Lett.\ {\bf 59} (1987) 26;
G.\ Gabrielse 
{\it et al.},
{\it ibid.},
{\bf 74} (1995) 3544.

\bibitem{bkr}
R.\ Bluhm, V.A.\ Kosteleck\'y and N.\ Russell,
Phys.\ Rev.\ Lett.\ {\bf 79} (1997) 1432;
Phys.\ Rev.\ D {\bf 57} (1998) 3932.

\bibitem{gg}
G.\ Gabrielse
{\it et al.},
in Ref.\ \cite{cpt98};
Phys.\ Rev.\ Lett.\ {\bf 82} (1999) 3198.

\bibitem{hd}
H.G.\ Dehmelt
{\it et al.},
Phys.\ Rev.\ Lett.\ {\bf 83} (1999) 4694.

\bibitem{rm}
R.K.\ Mittleman, I.I.\ Ioannou, and H.G.\ Dehmelt,
in Ref.\ \cite{cpt98};
R.K.\ Mittleman 
{\it et al.},
Phys.\ Rev.\ Lett.\ {\bf 83} (1999) 2116.

\bibitem {antih}
M.\ Charlton 
{\it et al.},
Phys.\ Rep.\ {\bf 241} (1994) 65;
J.\ Eades, ed., \it Antihydrogen, \rm
J.C.\ Baltzer, Geneva, 1993.

\bibitem{bkr2}
R.\ Bluhm, V.A.\ Kosteleck\'y and N.\ Russell,
Phys.\ Rev.\ Lett.\ {\bf 82} (1999) 2254.

\bibitem{bkl}
R.\ Bluhm, V.A.\ Kosteleck\'y and C.D.\ Lane, 
Phys.\ Rev.\ Lett., in press,
hep-ph/9912451.

\bibitem{cc}
V.W.\ Hughes, H.G.\ Robinson, and V.\ Beltran-Lopez,
Phys.\ Rev.\ Lett.\ {\bf 4} (1960) 342;
R.W.P.\ Drever,
Philos.\ Mag.\ {\bf 6} (1961) 683;
J.D.\ Prestage 
{\it et al.},
Phys.\ Rev.\ Lett.\ {\bf 54} (1985) 2387;
S.K.\ Lamoreaux 
{\it et al.},
{\it ibid.}, 
{\bf 57} (1986) 3125;
T.E.\ Chupp
{\it et al.},
{\it ibid.}, 
{\bf 63} (1989) 1541;
C.J.\ Berglund 
{\it et al.},
{\it ibid.}, 
{\bf 75} (1995) 1879.

\bibitem{kla}
V.A.\ Kosteleck\'y and C.D.\ Lane,
Phys.\ Rev.\ D {\bf 60} (1999) 116010;
J.\ Math.\ Phys.\ {\bf 40} (1999) 6245.

\bibitem{cfj}
S.M.\ Carroll, G.B.\ Field, and R.\ Jackiw,
Phys.\ Rev.\ D {\bf 41} (1990) 1231.

\bibitem{jk}
R.\ Jackiw and V.A.\ Kosteleck\'y,
Phys.\ Rev.\ Lett.\ {\bf 82} (1999) 3572.

\bibitem{pvc}
M.\ P\'erez-Victoria,
Phys.\ Rev.\ Lett.\ {\bf 83} (1999) 2518;
J.M.\ Chung,
Phys.\ Lett.\ B {\bf 461} (1999) 138.

\bibitem{bckp}
O.\ Bertolami
{\it et al.},
Phys.\ Lett.\ B {\bf 395} (1997) 178.

\bibitem{cg}
S.\ Coleman and S.\ Glashow,
Phys.\ Rev.\ D {\bf 59} (1999) 116008.

\bibitem{ra}
R.E.\ Allen,
these proceedings.

\bibitem{nm}
N.\ Mavromatos,
these proceedings.

\bibitem{dn}
D.\ Nanopoulos,
these proceedings.

\bibitem{hp}
H.\ P\"as,
these proceedings.

\bibitem{leewu}
T.D.\ Lee and C.S.\ Wu, 
Annu.\ Rev.\ Nucl.\ Sci.\ {\bf 16} (1966) 511.

\bibitem{hpk}
P.\ Haves and R.G.\ Conway,
Mon.\ Not.\ R.\ Astr.\ Soc.\ {\bf 173} (1975) 53P;
J.N.\ Clarke, P.P.\ Kronberg and M.\ Simard-Normandin,
{\it ibid.}, 
{\bf 190} (1980) 205.

\bibitem{ads}
A.D. Sakharov,
JETP Lett. {\bf 5} (1967) 24.

\end{thebibliography}
\end{document}